\pdfoutput=1
\documentclass[11pt,twoside,a4paper,cmspaper,final,collab]{cms-tdr}
\def\svnVersion{0745cfc}\def\svnDate{2026/11/24}\def\cmsCernNoTag{CERN-EP-2026-195}\def\cmsCernDate{\today}\def\cmsMessage{Published in Physics Letters B as \href{http://dx.doi.org/10.1016/j.physletb.2026.141105}{\doi{10.1016/j.physletb.2026.141105}.}}
\begin{document}\cmsNoteHeader{HIG-24-007}

\newlength\cmsFigWidth
\ifthenelse{\boolean{cms@external}}{\setlength\cmsFigWidth{\columnwidth}}{\setlength\cmsFigWidth{0.55\textwidth}}
\ifthenelse{\boolean{cms@external}}{\providecommand{\cmsLeft}{upper\xspace}}{\providecommand{\cmsLeft}{left\xspace}}
\ifthenelse{\boolean{cms@external}}{\providecommand{\cmsRight}{lower\xspace}}{\providecommand{\cmsRight}{right\xspace}}
\ifthenelse{\boolean{cms@external}}{\providecommand{\cmsTable}[1]{\resizebox{\columnwidth}{!}{#1}\xspace}}{\providecommand{\cmsTable}[1]{#1}}
\cmsNoteHeader{HIG-24-007}

\title{A measurement of the Higgs boson mass in the diphoton decay channel in proton-proton collisions at \texorpdfstring{$\sqrt{s} = 13\TeV$}{sqrt(s) = 13 TeV}}

\date{\today}
\newcommand{\Hgammagamma}{\ensuremath{\PH\to \PGg\PGg}\xspace}
\newcommand{\gammah}{\ensuremath{\Gamma_{\PH}}\xspace}
\newcommand*{\mgg}{\ensuremath{m_{\PGg\PGg}}\xspace}
\newcommand{\Ztoee}{\ensuremath{\PZ\to\Pe\Pe}\xspace}
\newcommand{\Ztomumu}{\ensuremath{\PZ\to\PGm\PGm}\xspace}
\newcommand{\Ztomumug}{\ensuremath{\PZ\to\PGm\PGm\PGg}\xspace}
\newcommand{\hgg}{\ensuremath{\PH\to\PGg\PGg}\xspace}
\newcommand{\hzz}{\ensuremath{\PH\to \PZ\PZ\to 4\Pell}\xspace}
\newcommand{\zz}{\ensuremath{\PZ\PZ\to 4\Pell}\xspace}
\newcommand{\mH}{\ensuremath{m_{\PH}}\xspace}
\newcommand{\Rn}{\ensuremath{R_\mathrm{9}}\xspace}
\newcommand{\Et}{\ensuremath{E_\mathrm{T}}\xspace}

\abstract{
A measurement of the Higgs boson mass in the diphoton decay channel is performed using proton-proton collision data at a center-of-mass energy of 13\TeV. The data set recorded with the CMS detector between 2016 and 2018 is used, corresponding to an integrated luminosity of 138\fbinv. A refined detector calibration and new analysis techniques are employed to improve the precision of the results compared to earlier measurements. The Higgs boson mass is measured to be $\mH = 125.13 \pm 0.15\GeV = 125.13 \pm 0.10\stat \pm 0.12\syst\GeV$. In addition, a combination with the mass measurement at center-of-mass energies of 7 and 8\TeV in the diphoton final state is performed resulting in $\mH = 125.06 \pm 0.14\GeV = 125.06 \pm 0.09\stat \pm 0.11\syst\GeV$. 
}

\hypersetup{%
pdfauthor={CMS Collaboration},%
pdftitle={A measurement of the Higgs boson mass in the diphoton decay channel in proton-proton collisions at sqrt(s) = 13 TeV},%
pdfsubject={CMS},%
pdfkeywords={CMS, Higgs boson, mass}}

\maketitle 

\section{Introduction}
\label{sec:intro}

The discovery of the Higgs boson ($\PH$) in 2012 by the ATLAS and CMS Collaborations in proton-proton collisions at the CERN LHC~\cite{20121, 201230, CMS:2013btf} was a breakthrough in the understanding of spontaneous symmetry breaking in the electroweak (EW) sector of the standard model of particle physics (SM). With the increase in the center-of-mass energy to $\sqrt{s}=13\TeV$ and the integrated luminosity achieved during the 2016--2018 LHC data-taking period, resulting in a substantially larger Higgs boson yield, the focus on precision measurements of the properties of the Higgs boson has become one of the cornerstones of the ATLAS and CMS Collaborations' physics programs. The Higgs boson mass, \mH, is not predicted by the SM, and precision measurements of \mH are further motivated by its connection with the nature of the stability of the EW vacuum, with profound cosmological implications~\cite{Markkanen:2018pdo}. Additionally, \mH is an input parameter to global fits of precision EW observables~\cite{ERLER201968}. Loop-level contributions to the $\PW$ boson propagator, involving the top quark and the Higgs boson, generate radiative corrections to the $\PW$ boson mass in the form:
\begin{equation*}
m_{\PW} = \frac{\Bigl(\frac{\pi \alpha}{G_{\text{F}} \sqrt{2}}\Bigr)^{\frac{1}{2}}}{\sin \theta_{\text{W}}} f(m_{\PQt}, \log(\mH)),
\end{equation*}
where $\alpha$ is the fine-structure constant, $G_{\text{F}}$ is the Fermi constant, $\theta_{\PW}$ is the Weinberg angle, and $f$ is a function of the top quark and Higgs boson masses~\cite{sirlin}; hence, precise measurements of the relevant parameters provide an over-constrained consistency test of the SM.

Even though the branching fraction of the diphoton decay channel \hgg is only 0.23\% for $\mH = 125\GeV$~\cite{deFlorian:2016spz}, it provides a fully reconstructable final state with excellent invariant-mass resolution, allowing a narrow signal peak to be distinguished over the continuum background. For this reason, the \hgg decay and the other high-resolution mode, \hzz ($\Pell \equiv \Pe,\PGm$), are the two best channels for a precision measurement of the Higgs boson mass.

The ATLAS Collaboration, using data collected at $\sqrt{s} = 13\TeV$ during 2015--2018 corresponding to an integrated luminosity of 140\fbinv, reported $\mH = 125.17 \pm 0.11\stat \pm 0.09\syst\GeV$  and $\mH = 124.99 \pm 0.18\stat \pm 0.04\syst\GeV$ in the $\PGg\PGg$ and \zz decay channels, respectively. These measurements, combined with earlier measurements of \mH at $\sqrt{s} = 7$ and $8\TeV$, led to the observed value of $\mH = 125.11 \pm 0.09\stat \pm 0.06\syst\GeV$~\cite{PhysRevLett.131.251802}.

The CMS Collaboration measured \mH in the diphoton decay channel using LHC data collected  at $\sqrt{s} = 7$ and $8\TeV$ during 2011--2012 and at $\sqrt{s} = 13\TeV$ during 2016. The measured values were $\mH = 124.70 \pm 0.31\stat \pm 0.15\syst\GeV$~\cite{Run1HGG} and $\mH = 125.78 \pm 0.18\stat \pm 0.18\syst\GeV$~\cite{higgs_mass}, respectively. These measurements were combined with those from the \zz decay channel using data collected during the same periods, yielding a combined mass measurement of $\mH = 125.38 \pm 0.11\stat \pm 0.08\syst\GeV$~\cite{higgs_mass}. Subsequently, the analysis in the \zz decay channel was extended to include the full data set collected during 2016--2018 at $\sqrt{s} = 13\TeV$, resulting in a measured value of $\mH = 125.04 \pm 0.11\stat \pm 0.05\syst\GeV$~\cite{CMS:2023hig4l}.

In this Letter, we present a measurement of the Higgs boson mass in the \hgg decay channel using data collected at $\sqrt{s} = 13\TeV$ between 2016 and 2018, corresponding to an integrated luminosity of 138\fbinv, and its combination with the measurement performed at $\sqrt{s} = 7$ and $8\TeV$. Special emphasis is placed on the methods used to mitigate and quantify the main sources of systematic uncertainty. In this measurement, the electron energy scale is measured using \Ztoee decays, and the photon-to-electron energy scale ratio is determined with photons from \Ztomumug decays. The combination of these two techniques removes uncertainties related to the differences between electrons and photons in the calibration.

\section{The CMS detector}
\label{sec:detector}
The CMS apparatus~\cite{CMS:2008xjf,CMS:2023gfb} is a multipurpose, nearly hermetic detector, designed to trigger on~\cite{CMS:2020cmk,CMS:2016ngn,CMS:2024aqx} and identify electrons, muons, photons, and charged and neutral hadrons~\cite{CMS:2020uim,CMS:2018rym,CMS:2014pgm}. Its central feature is a superconducting solenoid of 6\unit{m} internal diameter, providing a magnetic field of 3.8\unit{T}. Within the solenoid volume are a silicon pixel and strip tracker, a lead tungstate crystal electromagnetic calorimeter (ECAL), and a brass and scintillator hadron calorimeter (HCAL), each composed of a barrel and two endcap sections. The ECAL is a hermetic homogeneous calorimeter made of $61\,200$ lead tungstate ($\text{PbWO}_4$) crystals mounted in the central barrel part, closed by 7324 crystals in each of the two endcaps. In the region $1.65 < \abs{\eta} < 2.60$ a three radiation length thick preshower detector with two orthogonal layers of silicon strips is placed in front of the endcap crystals. Avalanche photodiodes are used as the photodetectors in the barrel and vacuum phototriodes in the endcaps. The barrel part of the ECAL covers the pseudorapidity range  $\abs{\eta} < 1.48$, while the endcap calorimeters cover the range $1.48 < \abs{\eta} < 3.00$. The electronic readout of each crystal has three amplifiers with different gains to cover the energy range from 30\MeV to 1\TeV. Muons are reconstructed using gas-ionization detectors interleaved with the layers of the steel flux-return yoke outside the solenoid. The first level of the CMS trigger system uses information from the calorimeters and muon detectors to select the most interesting events in a fixed time interval of about 4 $\mu$s. The high-level trigger, implemented on a processor farm, further decreases the event rate from around 100\unit{kHz} to a few \unit{kHz} before data storage. A more detailed description of the CMS detector, together with a definition of the coordinate system used and the relevant kinematic variables, can be found in Refs.~\cite{CMS:2008xjf,CMS:2023gfb}.

\section{Analysis strategy}
\label{sec:strat}

In this Letter, we present a measurement of the Higgs boson mass \mH following a strategy similar to that used in Ref.~\cite{higgs_mass}. One diphoton pair per event is selected and a score is assigned to each pair based on a discriminant trained to separate the \hgg signal from the continuum diphoton background originating mainly from prompt photons and jets. The selected diphoton pairs are then categorized and a simultaneous fit to the \mgg distribution in all categories is performed to extract the value of \mH.

Several improvements with respect to Ref.~\cite{higgs_mass} have been introduced to reduce the dominant sources of systematic uncertainty, in particular those related to the measurement of the photon energy. A refined calibration is performed to improve the energy scale stability and the energy resolution of the ECAL detector~\cite{Hayrapetyan_2024}, affecting the calorimetric deposits used to reconstruct the energies of electrons and photons. The effect of the interference between the gluon-fusion-induced \hgg signal process, $\Pg\Pg\PH$, and the continuum diphoton background~\cite{hgginterference1, hgginterference2} is taken into account as a systematic uncertainty.

In Ref.~\cite{higgs_mass}, the ECAL energy response was calibrated using electrons from \Ztoee decays. The photon energy scale was extrapolated from the electron energy scale using a \GEANTfour~\cite{Agostinelli:2002hh, Allison:2006ve, Allison:2016lfl}-based simulation of the CMS detector. This procedure accounted for the effect of radiation damage on the transparency of the ECAL crystals, that affects the energy scales of electrons and photons differently, and was the dominant source of systematic uncertainty.

In this work, we have adopted an improved energy calibration strategy that builds on the methods described in Ref.~\cite{IJazZ}. The electron energy scales are first measured using \Ztoee decays, then the radiation-induced differences between the electron and photon energy scales are accounted for using corrections derived from simulation. The residual differences between the photon and electron energy scales are then measured directly using photons from \Ztomumug decays. To further improve the precision of the measurement, a multivariate signal-to-background discriminator is employed, with the background modeled using control samples in data. The event categorization is refined using the signal significance, defined as $S/\sqrt{B}$, where $S$ and $B$ are the numbers of signal and background events, respectively, in a window around the signal peak, together with the expected diphoton mass resolution.

\section{Data and simulation}
\label{sec:samples}

The proton-proton collisions data sample used in this analysis was collected between 2016 and 2018 at the LHC with an integrated luminosity of 138\fbinv~\cite{CMS-LUM-17-003, CMS-PAS-LUM-17-004, CMS-PAS-LUM-18-002} at a center-of-mass energy of 13\TeV.
During the 2016 data taking period, events were selected with a diphoton high-level trigger with asymmetric transverse momentum (\pt) thresholds on the photons of 30 and 18\GeV. For the data collected during 2017 and 2018 the threshold for the lower \pt photon was increased to 22\GeV to accommodate the higher instantaneous luminosity. Details of the trigger selection and the measurement of the trigger efficiency can be found in Ref.~\cite{CMS:2021kom}.

The gluon-gluon fusion, vector-boson fusion, associated vector-boson, and associated top-quark pair Higgs boson production modes are considered in the analysis. All the Higgs boson signal production modes are simulated with the \MGvATNLO 2.6.5 matrix-element Monte Carlo (MC) event generator~\cite{AMCAT} at next-to-leading order (NLO) in quantum chromodynamics (QCD), interfaced with \PYTHIA 8.240~\cite{SJOSTRAND2015159} for parton showering and hadronization. The FxFx jet merging scheme at NLO~\cite{Frederix:2012ps} is used in all cases except for associated vector-boson production, while the \PYTHIA underlying event tune CP5~\cite{CMS:2019csb} is used in all cases. All signal samples are normalized to the cross sections provided by the LHC Higgs Working Group for a center-of-mass energy of 13\TeV~\cite{deFlorian:2016spz}.

In addition, the interference between the gluon-fusion-induced \hgg signal process and the continuum diphoton background process is simulated with the \SHERPA 2.2.5~\cite{SHERPA2} generator and is used to estimate the shift in the peak position of the Higgs boson lineshape. The estimated magnitude of the mass shift due to the interference mechanism is treated as a systematic uncertainty, as discussed in Section \ref{sec:systematics}. The \SHERPA generator is also used to simulate the irreducible diphoton background, which includes the Born processes with up to three additional jets at leading-order (LO) accuracy in perturbative QCD, as well as the LO box processes.

The $\PGg$+jets and multijet backgrounds, used to train a discriminant to distinguish between signal and background events, are modeled with a technique based on control samples in data discussed in Section \ref{sec:categories}. In the background estimation, $\PGg$+jets samples, simulated with \PYTHIA at LO, are used to model the distribution of the identification (ID) values for misidentified jets passing the photon ID requirement. In the production of the $\PGg$+jets samples, a filter is applied to select preferentially jets with a large fraction of energy coming from electrons or photons. The event filter is applied at the generator level, requiring at least two electrons or photons with $\pt > 15\GeV$, while the invariant mass of the pair must exceed $80\GeV$.

The electron and photon energy scale corrections, the data-to-simulation correction factors related to the photon selections, and the related systematic uncertainties were derived using Drell--Yan samples simulated with \MGvATNLO at NLO precision, interfaced with \PYTHIA for parton showering and hadronization via FxFx jet merging scheme at NLO~\cite{Frederix:2012ps}. The CP5 underlying event tune was used.

All simulated samples include multiple proton-proton interactions taking place in the same or in neighboring bunch crossings (pileup). The average number of these pileup events in the simulation is adjusted to reproduce their observed distribution in the collected data. All of the simulated events are propagated through the full CMS detector simulation using the \GEANTfour package~\cite{Agostinelli:2002hh}.

\section{Photon reconstruction and identification}
\label{sec:phoRecoAndID}

The photon candidates are reconstructed following the same procedure as in Refs.~\cite{HIG-16-040,CMS:2020uim}, which includes the particle-flow algorithm~\cite{CMS-PRF-14-001}, where cluster `seed' crystals are identified as local energy maxima in the ECAL above a given threshold, then clusters are grown around the seeds by aggregating crystals that share an edge with the seed when their energy is above a certain threshold. The energy of each crystal in adjacent clusters is shared assuming a Gaussian transverse profile of the electromagnetic shower. Finally, clusters are merged into extended clusters, or groups of clusters known as `superclusters' with a procedure that accounts for variations in geometry as a function of $\abs{\eta}$, and optimizes the robustness of the energy resolution against pileup. Superclusters are constructed to recover the energy of secondary photons emitted through electron bremsstrahlung and of electrons produced in photon conversions.

\subsection{Photon energy calibration}

A critical component of the measurement of the Higgs boson mass in this decay channel is the photon energy measurement by the ECAL.
To calculate the energy of a photon, the energies of all the crystals associated with the supercluster are summed~\cite{Hayrapetyan_2024}, and in the region $1.65 < \abs{\eta} < 2.60$ the energy deposited in the preshower is also included. The energy scale is corrected to account for incomplete containment of the electromagnetic showers in the supercluster, energy lost by photons that convert upstream of the calorimeter, and the effect of pileup. These corrections are derived using a multivariate regression technique, trained on simulated events, that estimates the energy of the photon and its uncertainty~\cite{CMS:2020uim}. The inputs to this regression are variables that describe the shape of the electromagnetic shower, the fraction of energy deposited in the preshower, and observables sensitive to pileup. To ensure that the simulation accurately describes the data, the differences between the data and simulation in both the photon energy scale and the energy resolution are corrected for with a three-stage calibration procedure. In the first stage, electrons from \Ztoee events are used to derive energy scale and resolution corrections; in the second stage, differences in the energy scale between electrons and photons are corrected using the simulation; and in the third stage, final-state-radiation (FSR) photons coming from \Ztomumug events are used to correct for residual differences in the energy scale between photons and electrons in data.

To account for differences between data and simulation, a reweighting procedure is applied in the first and third stages of the calibration, which corrects for the residual disagreements in the transverse momentum distribution of the Z boson candidates, in the $\eta$ distribution of the two electrons in the first stage, or of the single photon in the third stage, and in the pileup distribution, estimated using the median energy density per unit area in the transverse plane.

The first stage of the calibration uses \Ztoee events from data and simulation following the method described in Ref.~\cite{IJazZ}, which maximizes a likelihood function sensitive to the energy scale and resolution, after reconstructing the electron energy as if it were a photon. For every data taking period, energy scale corrections are derived from the \Ztoee sample and applied to every electron. These corrections are derived through five iterative steps, described below, with the final correction computed as the product of the corrections at each iteration, and applied as a multiplicative factor to the particle energy in data. In addition, the energy resolution in simulated events is matched to that observed in data by applying a random Gaussian broadening to the electron energy, the magnitude of which is at most 2\%.

For each supercluster, a shower shape variable \Rn is defined, which is the ratio of the sum of the energy deposited in the $3{\times}3$ crystal array, with the highest energy crystal at the center, to the sum of the energy in the whole supercluster. The energy deposits of photons that interact before reaching the calorimeter tend to have a wider transverse profile and, thus, lower values of \Rn than those of unconverted photons.

In the first step, drifts in the long-term energy scale of the data, mostly due to irradiation, are corrected in bins of $\eta$ in time intervals corresponding to roughly one LHC fill~\cite{CMS-LUM-17-003}.
In the second step, corrections are derived simultaneously in bins of $\abs{\eta}$ and \Rn.
In the third step, the corrections are derived in bins of \Rn, $\abs{\eta}$, and \Et, where $\Et = E/\cosh{\eta}$ is the transverse energy, to account for any non-linearity in the energy response for electrons. 
In the fourth step, the corrections are derived separately for the three electronics gain ranges used to read out the seed crystal of the supercluster. Finally, although the energy scale and resolution corrections are extracted simultaneously at each step, the resolution corrections and corresponding systematic uncertainties are estimated using \Et bins twice as wide as in the previous steps. Compared to the measurement in Ref.~\cite{higgs_mass}, the energy resolution corrections are \Et-dependent, in addition to being derived in bins of \Rn and $\eta$.

This fine-grained \Et-dependent energy scale correction improves the precision of the Higgs boson mass measurement, since the mean energy of the electrons in $\PZ$ boson decays (${\approx}45\GeV$) used to derive the scale corrections is lower than the mean energy of photons in Higgs boson decays (${\approx}60\GeV$) where these corrections are applied.

\begin{figure}[htb!]
    \centering
    \includegraphics[width=\cmsFigWidth]{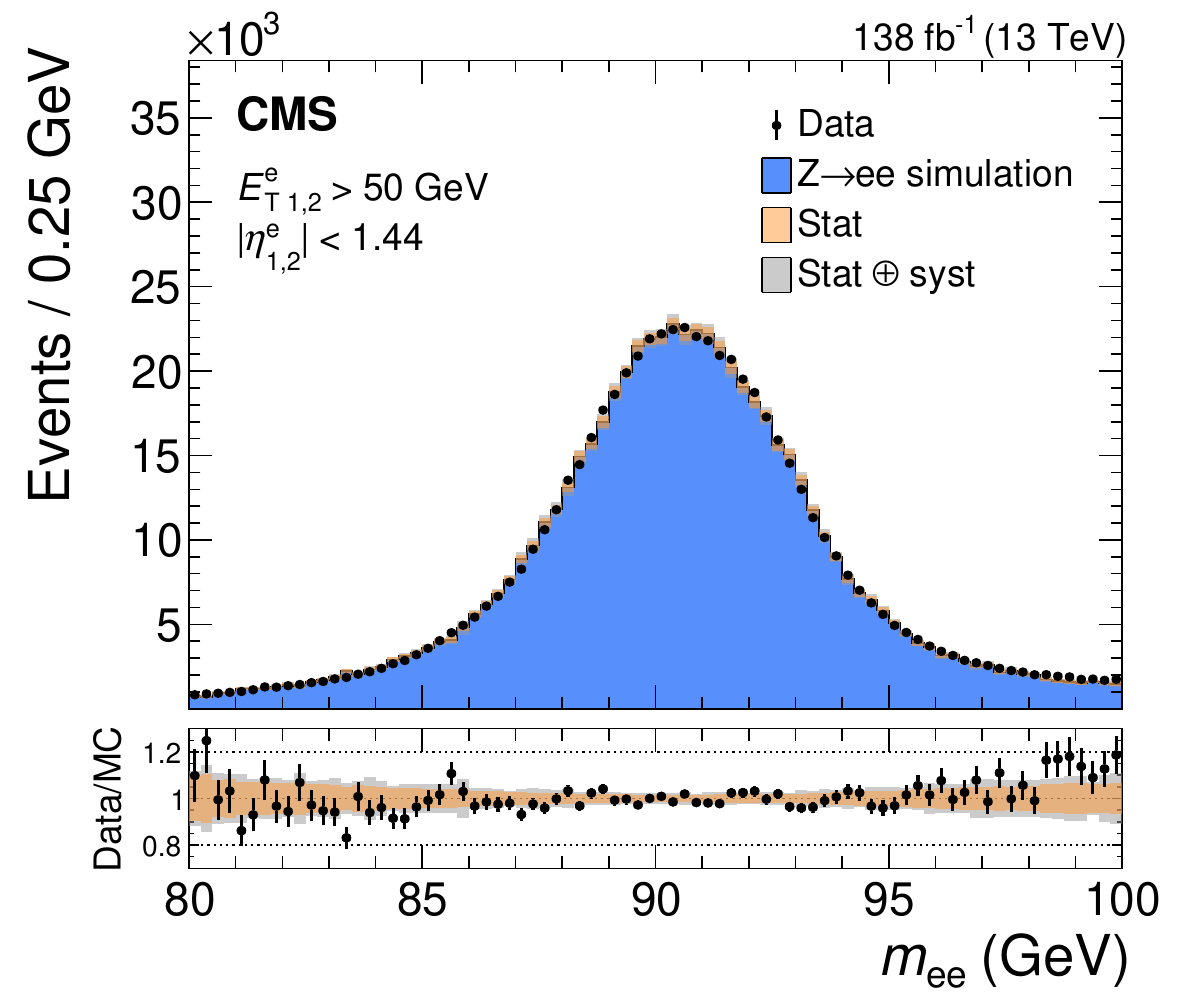}
    \caption{Comparison of the dielectron invariant mass distribution in data and simulation for \Ztoee events after applying only the electron energy corrections. Both electrons are required to have $E^{\Pe}_{\mathrm{T}} > 50\GeV$ and to be reconstructed in the ECAL barrel region. The lower panel shows the bin-by-bin ratio of the data to the simulation (MC). The error bands account for the statistical and systematic uncertainties associated with the energy scale corrections.}
    \label{fig:inv_mass_agreement}
\end{figure}

A comparison between data and simulation, after applying the energy scale corrections and the energy resolution broadening, can be seen in the dielectron invariant mass distribution shown in Fig.~\ref{fig:inv_mass_agreement}, demonstrating the performance of the energy scale corrections in the kinematic region where the \Et of the two electrons exceeds $50\GeV$, corresponding to the range relevant for photons from Higgs boson decays.

In the second stage, an additional correction factor, the `uniformity' correction, is introduced to account for the variation in time of the radiation damage sustained by the ECAL crystals, that affects the uniformity of light collection and consequently the energy scale. While the average radiation-induced loss in each crystal is corrected using a laser monitoring system~\cite{Hayrapetyan_2024}, residual differences between photons and electrons may arise because the radiation damage mainly affects the crystal's front face, making the light collection non-uniform. Electromagnetic showers initiated by photons develop approximately 0.8 radiation lengths (${\approx} 7$\unit{mm}) deeper in the crystal than showers initiated by electrons, leading to a difference in the energy response. The uniformity correction takes into account the increase in the radiation damage to the crystals, and any radiation-induced change in the response of the photodetectors. It is computed as the ratio of the electron and photon responses in the ECAL, where the response is defined as the convolution of the shower profile, simulated with \GEANTfour and the light collection efficiency. A \textsc{Fluka}~\cite{FLUKA} simulation is used to estimate the radiation damage and \textsc{Litrani}~\cite{LITRANI} is used as a ray-tracing simulation to estimate the non-uniformity of light collection. This correction is applied only to high-\Rn photons ($\Rn > 0.96$), since a low-\Rn photon ($\Rn \leq 0.96$) has a shower profile that is closer to that of an electron.

\begin{figure}[htb!]
\centering
\includegraphics[width=0.49\textwidth]{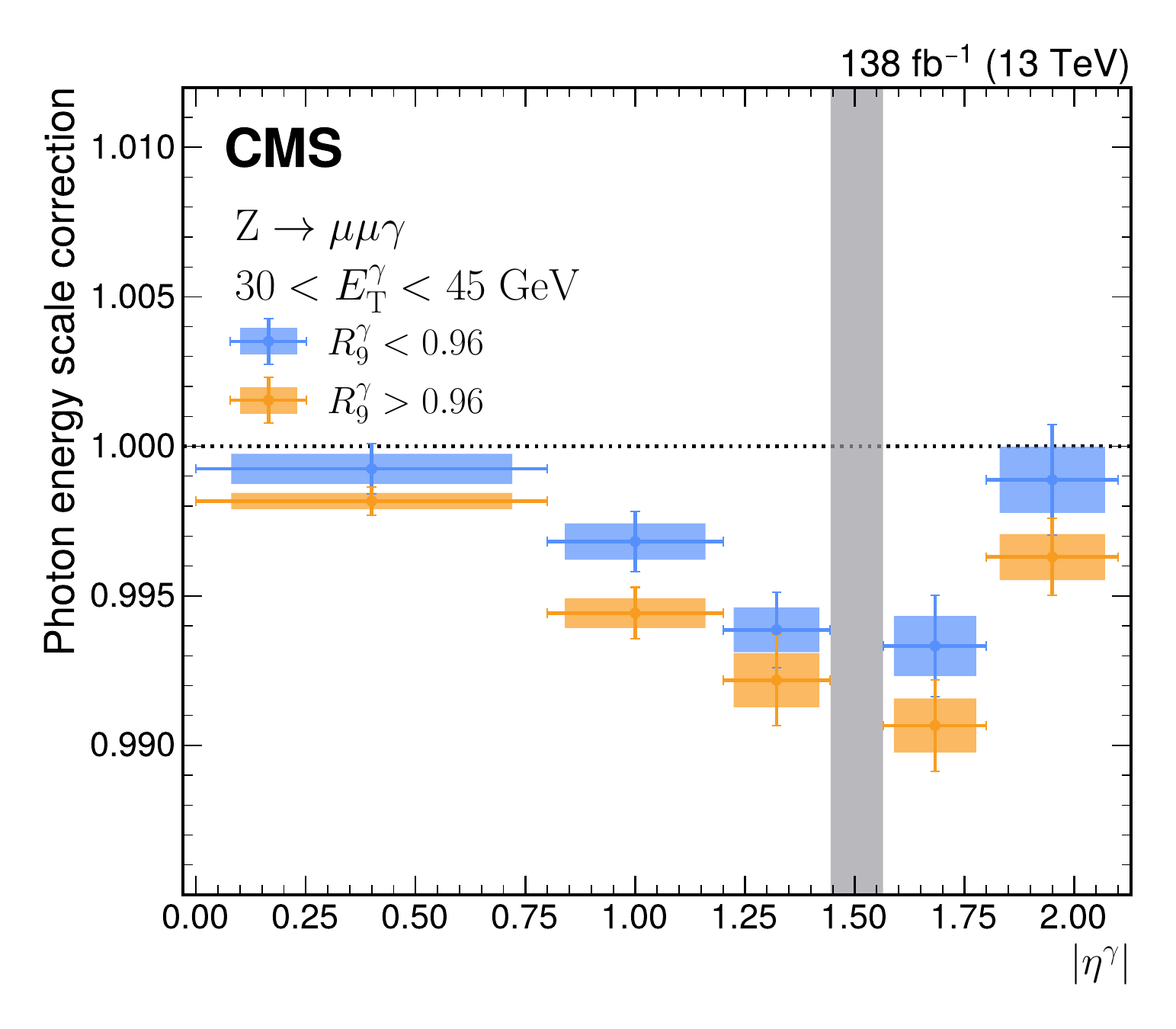}
\includegraphics[width=0.49\textwidth]{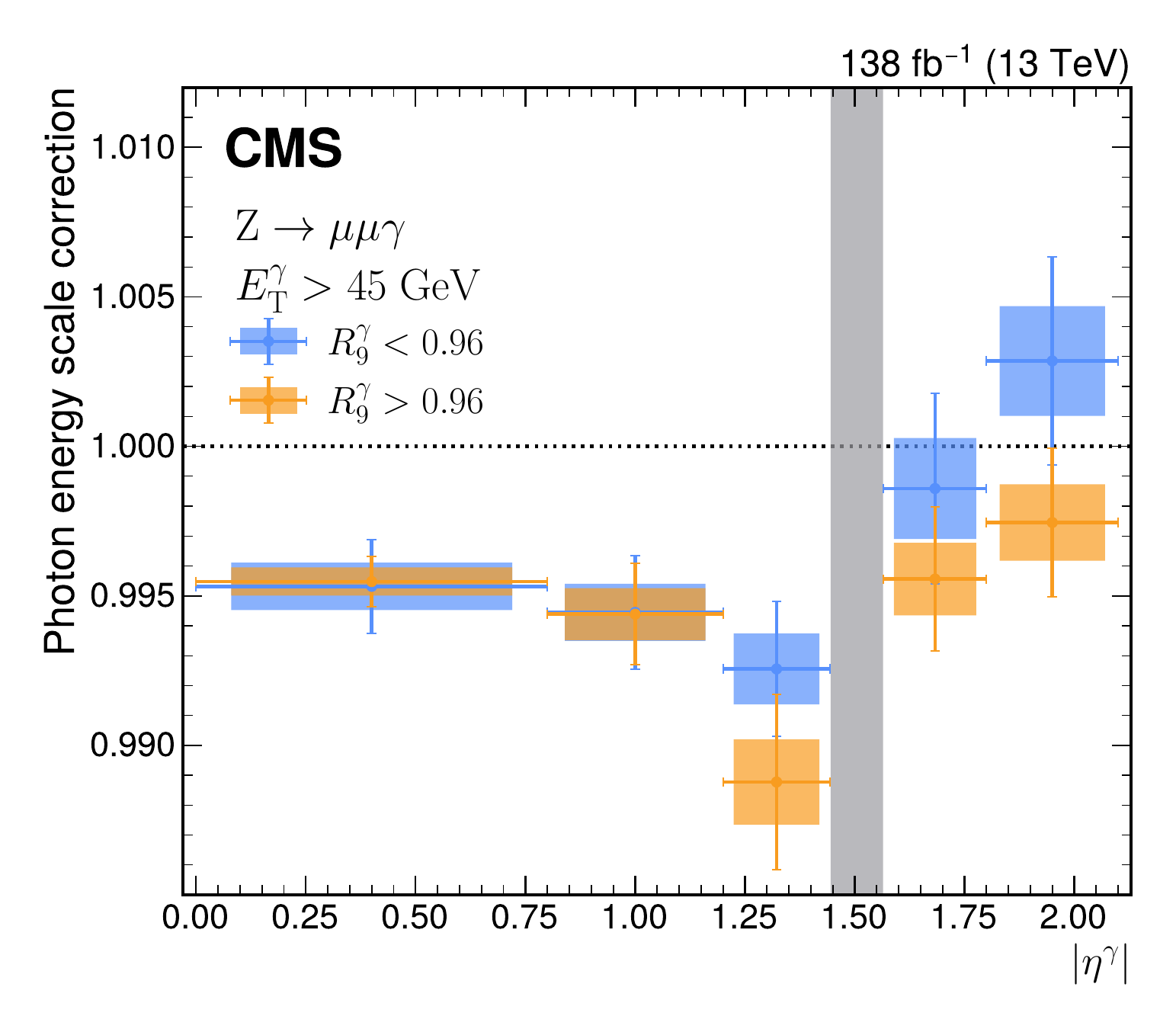}
\caption{Photon energy scale corrections applied to the data as functions of $\abs{\eta^\PGg}$ shown separately for the two $E^{\PGg}_{\mathrm{T}}$ bins and the two $R^{\PGg}_{\mathrm{9}}$ bins, using \Ztomumug events, after applying the electron energy scale and uniformity corrections. The shaded bands indicate the statistical uncertainties from the simulation, while the error bars show the total statistical uncertainties obtained by summing in quadrature the uncertainties from the data and the simulation. The gray vertical band represents the ECAL barrel-endcap transition region, which is not used in the analysis.}
\label{fig:ZmmgScale}
\end{figure}

Finally, in the third stage, the photon energy scale is further corrected using the FSR-photon from \Ztomumug events. The method used in the electron energy scale calibration~\cite{IJazZ}, based on the \Ztoee invariant mass distribution, was adapted to derive the final photon energy scale corrections. For this, a new observable sensitive to photon energy miscalibration in the three-body decay is defined and used instead of the dielectron invariant mass, and the analytical likelihood method adapted accordingly. Events are selected by requiring the presence of two muons with opposite charge satisfying tight identification criteria~\cite{CMS:2018rym}. The leading and subleading muons are required to have a transverse momentum exceeding 20 and 10\GeV respectively. In addition, the photon is required to have $\Et > 30\GeV$, to pass a loose requirement on the photon ID score, and to be separated from the nearest muon by $0.01 < \Delta R(\PGm,\PGg) < 0.8$, where $\Delta R(\PGm,\PGg) = \sqrt{\smash[b]{(\eta^{\PGm}-\eta^{\PGg})^2 + (\phi^{\PGm}-\phi^{\PGg})^2}}$ and $\phi$ is the azimuthal angle. Due to the small number of \Ztomumug events in data compared to \Ztoee, the full data sample is used to derive a correction common to all three data taking periods.
The photon energy scale corrections are determined in 20 bins: five bins in $\abs{\eta}$, two bins in \Rn (low and high), and two bins in \Et. The low-\Et bin is defined as $30 < \Et < 45\GeV$, while the high-\Et bin is defined as $\Et> 45\GeV$. The photon energy scale corrections applied to the data are shown in Fig.~\ref{fig:ZmmgScale}, where the correction factors range from 0.990 to 1.000. 

\begin{figure}[htb!]
\centering
\includegraphics[width=0.49\textwidth]{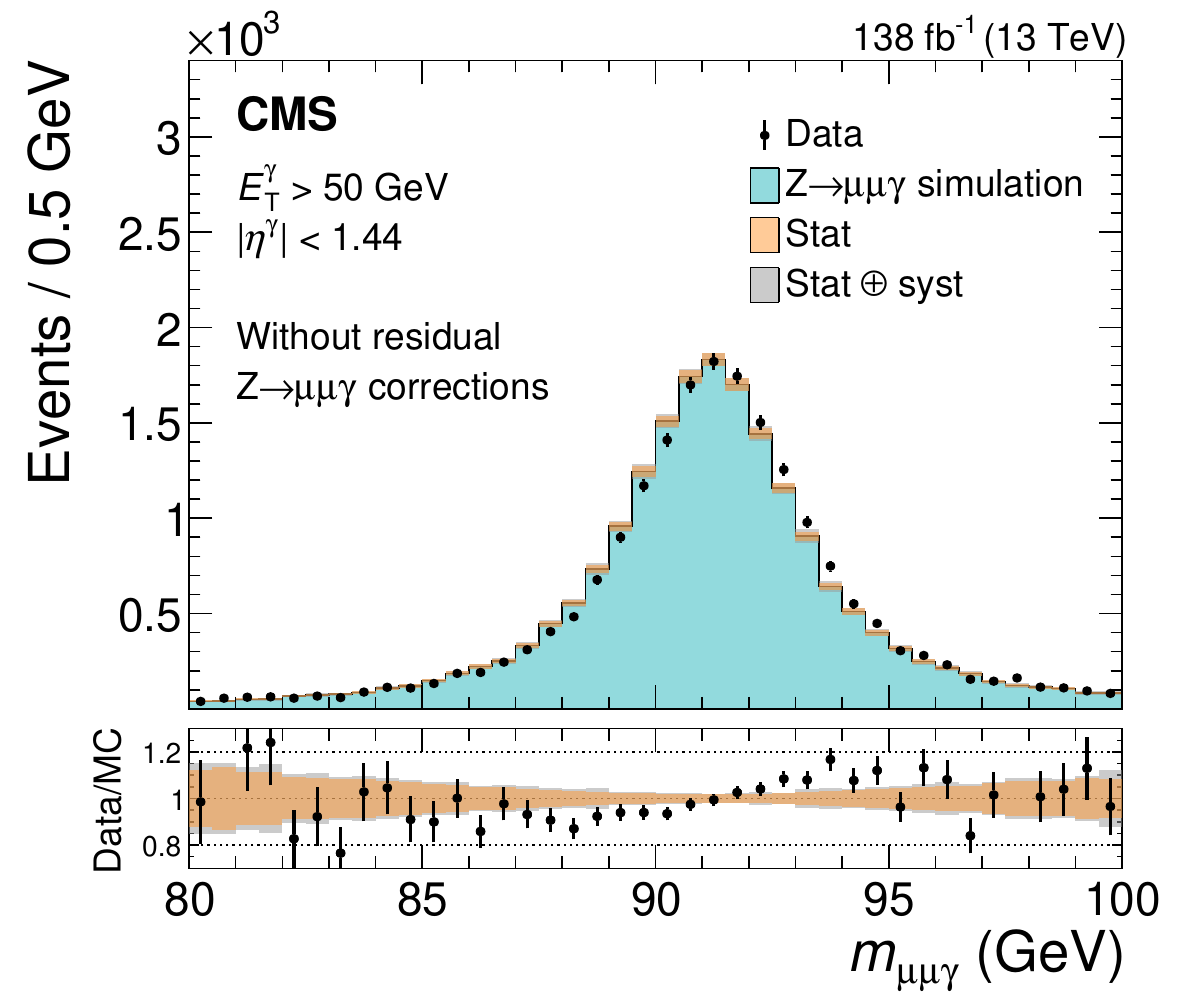}
\includegraphics[width=0.49\textwidth]{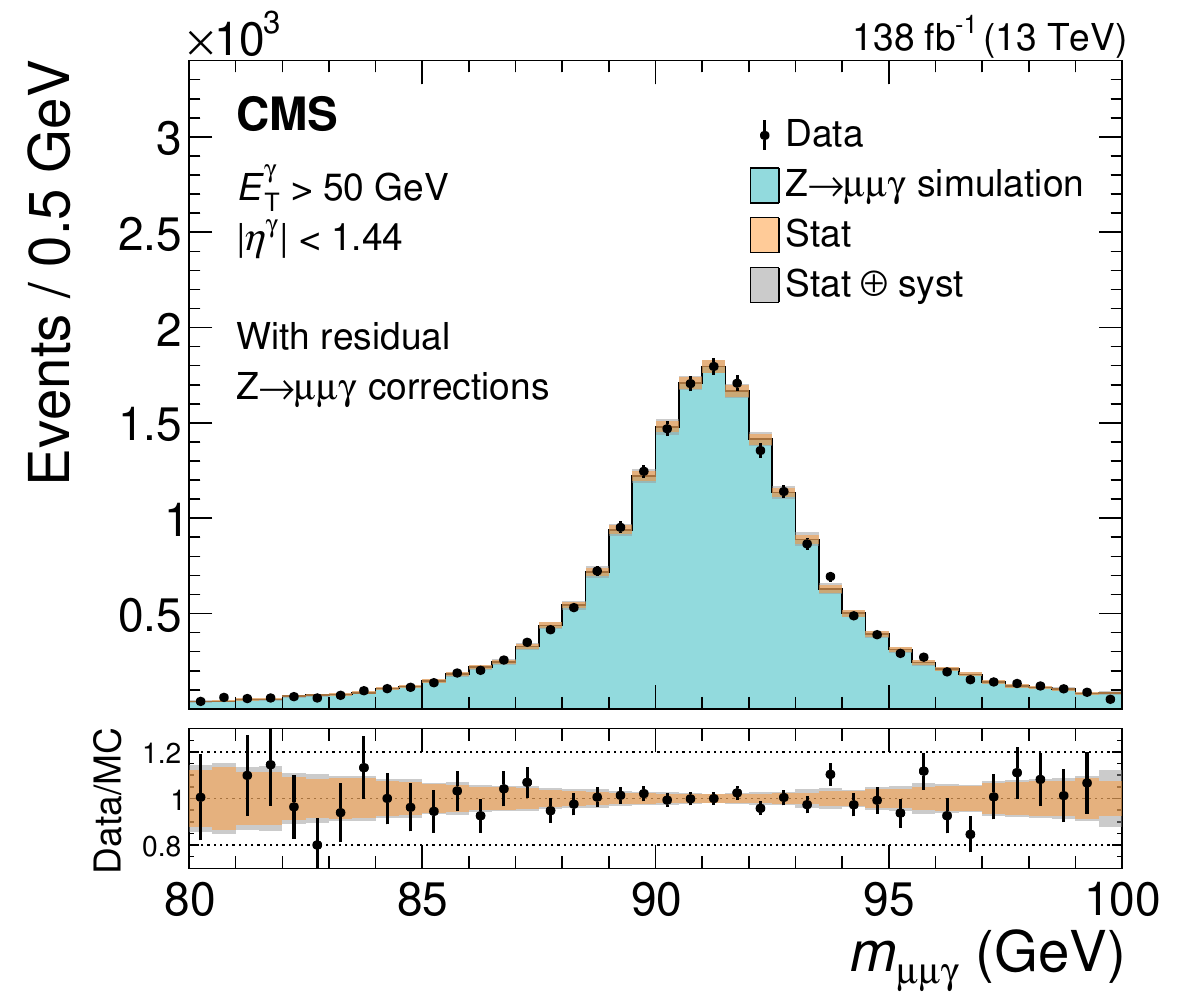}
\caption{Comparison of the three-body invariant mass distribution in data and simulation for \Ztomumug events in the ECAL barrel region ($\abs{\eta^\PGg} < 1.44$) with $E^{\PGg}_{\mathrm{T}} > 50\GeV$. The \cmsLeft panel shows the distribution after the first two calibration stages, without the final photon energy scale correction derived from FSR photons. The \cmsRight panel shows the same distribution after applying the final photon energy scale corrections. The lower panels show the bin-by-bin ratio of the data to the simulation (MC). The error bands account for the statistical and systematic uncertainties associated with the energy scale corrections.}
\label{fig:ZmmgVal}
\end{figure}

The effect of the final corrections is illustrated in Fig.~\ref{fig:ZmmgVal}, which compares the three-body invariant mass distribution in data and simulation before and after their application. Events are selected to have a photon with $\Et > 50\GeV$, corresponding to the range relevant for photons from Higgs boson decays. A discrepancy between data and simulation is observed before the application of the final corrections. Following their application, the data agree with the simulation within the statistical and systematic uncertainties.

\subsection{Photon preselection and identification}

The photons used in the subsequent steps of the analysis are required to satisfy preselection criteria that are similar to, but more stringent than, those imposed by the trigger. A detailed description of these criteria and the methods used to evaluate their efficiencies can be found in Ref.~\cite{HIG-16-040}. A boosted decision tree (BDT) classifier is then used to distinguish prompt photons from candidates arising from misidentified hadronic jets that satisfy the preselection criteria. Details of the BDT and its input features are also provided in Ref.~\cite{HIG-16-040}.

\subsection{Vertex selection}

The event-by-event uncertainty in the reconstructed diphoton invariant mass depends on the uncertainties in the photon energies and their opening angle. When the selected diphoton vertex is within 10\unit{mm} of the true vertex, the contribution from the angular term is subdominant compared to that from the photon energy resolution. The diphoton vertex is chosen following the procedure described in Ref.~\cite{HIG-16-040}. A BDT, using observables related to tracks recoiling against the diphoton system, identifies the most likely vertex, while a second BDT estimates the probability that this vertex is correct. The algorithm is validated using \Ztomumu events with the muon tracks removed to mimic diphoton production. In simulated \hgg events, the efficiency of assigning the event to a vertex within 10\unit{mm} of the true vertex depends on the diphoton transverse momentum and is found to be 79\% on average.

\section{Event classification}
\label{sec:categories}

The event selection procedure is similar to that in Ref.~\cite{higgs_mass}. The \pt values of the two leading photons ($p^{\PGg}_{\textrm{T},1}$, $p^{\PGg}_{\textrm{T},2}$) are required to satisfy $p^{\PGg}_{\textrm{T},1} > \mgg/3$ and $p^{\PGg}_{\textrm{T},2} > \mgg/4$, where the photon \pt requirement is applied after the vertex assignment and \mgg is the invariant mass of the selected diphoton pair. The \mgg scaled \pt thresholds are used to reduce a distortion at the lower end of the invariant mass spectrum due to trigger thresholds, which would otherwise affect the background modeling. Additionally, \mgg is required to be between 100 and 180\GeV, and the superclusters of both photons are required to have a pseudorapidity $\abs{\eta} < 2.10$, and to be outside of the barrel-endcap transition region, $1.44 < \abs{\eta} < 1.57$.

To maximize the discrimination of \hgg signal events from the irreducible diphoton continuum and reducible backgrounds, a diphoton BDT classifier trained with the \textsc{XGBoost} framework~\cite{Chen:2016btl} is used. A high score is assigned to events with diphoton pairs exhibiting signal-like kinematic properties. The input variables to the classifier~\cite{HIG-16-040} are: $\pt^{\PGg}/\mgg$ for each photon, the pseudorapidity of the two photons, the cosine of the angle between the two photons in the transverse plane, the photon ID scores for each photon, the per-event relative diphoton mass resolution computed from the photon energy resolution and the uncertainty in the diphoton opening angle due to the length of the beam spot, and the per-event probability estimate that the correct primary vertex has been assigned to the diphoton candidate. This set of input variables is chosen to ensure that selections on the classifier output preserve the smoothly falling behavior of the diphoton mass spectrum, confirmed in dedicated studies. The diphoton BDT is trained with all the Higgs boson production modes as the positive class against SM diphoton backgrounds as the negative class. 
To improve the classifier performance, some of the backgrounds used during training are estimated using a large sample of data events. In particular, events where at least one photon candidate fails the photon ID requirement are used to model the backgrounds with at least one jet misidentified as a photon. This sample is almost exclusively composed of multijet events and events with a prompt photon accompanied by jets. More details on this background estimation technique based on control samples in data can be found in Ref.~\cite{tthprl}. A sample of simulated prompt diphoton background events is also used to train the BDT. The modeling of the input features and the performance of the classifier have been validated in signal-depleted diphoton events and in \Ztoee events.

For each data taking period, the events are first sorted into two groups: one with both photons in the ECAL barrel and the other with at least one photon in the ECAL endcaps, with the first group containing events with a better expected mass resolution. The first group is further subdivided into categories defined by bins of the diphoton BDT classifier score. The bin boundaries are chosen to minimize the relative diphoton invariant mass resolution while simultaneously maximizing the statistical significance in each bin. A total of four analysis categories with both photons in the ECAL barrel and one category with at least one photon in the ECAL endcaps are defined. This analysis uses data sets collected between 2016 and 2018. In 2016, due to changes in data-taking conditions, the data were divided into two independent samples. This results in a total of 20 signal categories, which after optimization have a statistical significance between 0.6 and 1.8.

\section{Statistical analysis}
\label{sec:statAnalysis}

To extract the mass of the Higgs boson, signal and background models are constructed to fit the diphoton mass distributions observed in data.
Since the \mgg resolution depends on the properties of the photons, and as a consequence on the production mechanism and the analysis category, a signal model is derived for each production mechanism and analysis category using simulated Higgs boson events, while the background model is derived directly from data. To determine the observed value of \mH and its uncertainty, a simultaneous binned maximum likelihood fit is performed to the \mgg distributions of all analysis categories in the range $100 < \mgg <  180\GeV$. The dependence of the results on the bin size has been tested and found to be negligible. 

\subsection{Signal model}
\label{sec:results_signal}

The simulated events used to derive the signal shapes for each analysis category and production mode are generated assuming a nominal value of $\mH=125\GeV$, taking into account the trigger, reconstruction, and identification efficiencies. For each process, analysis category, and data taking period, the simulated $\mgg$ distribution is fitted with a sum of at most five Voigtian functions, where the Voigtian is the convolution of a Breit--Wigner distribution and a Gaussian distribution~\cite{Voigt1912}, and the resonance width parameter, \gammah, of each Voigtian is set to the SM expectation of 4.1\MeV for a 125\GeV Higgs boson \cite{deFlorian:2016spz}. The number of Voigtian components is chosen to minimize the $\chi^2/\textrm{dof}$ with respect to the simulated \mgg distribution. 

\begin{figure}[htb!]
  \centering
  \includegraphics[width=0.49\textwidth]{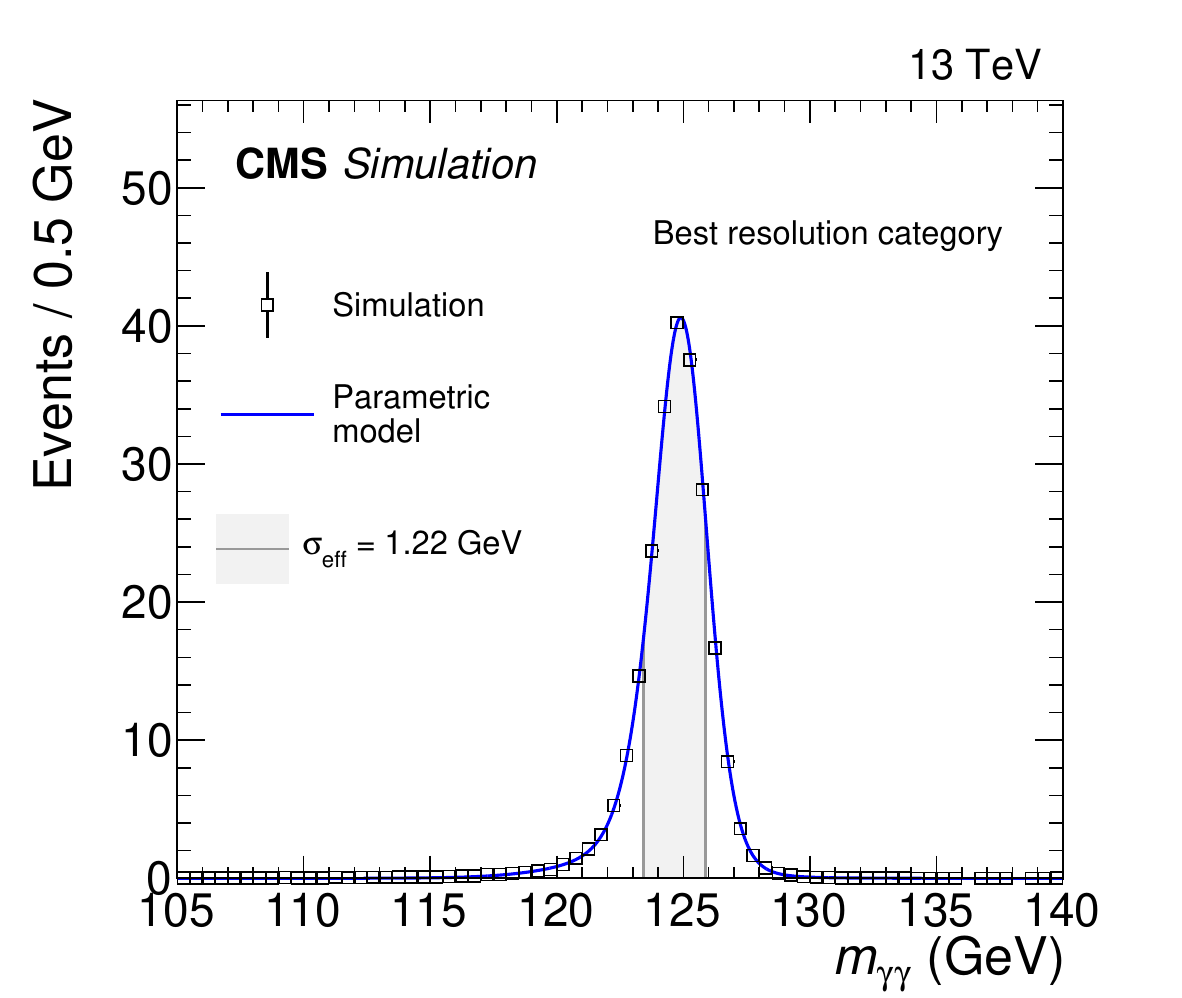}
  \includegraphics[width=0.49\textwidth]{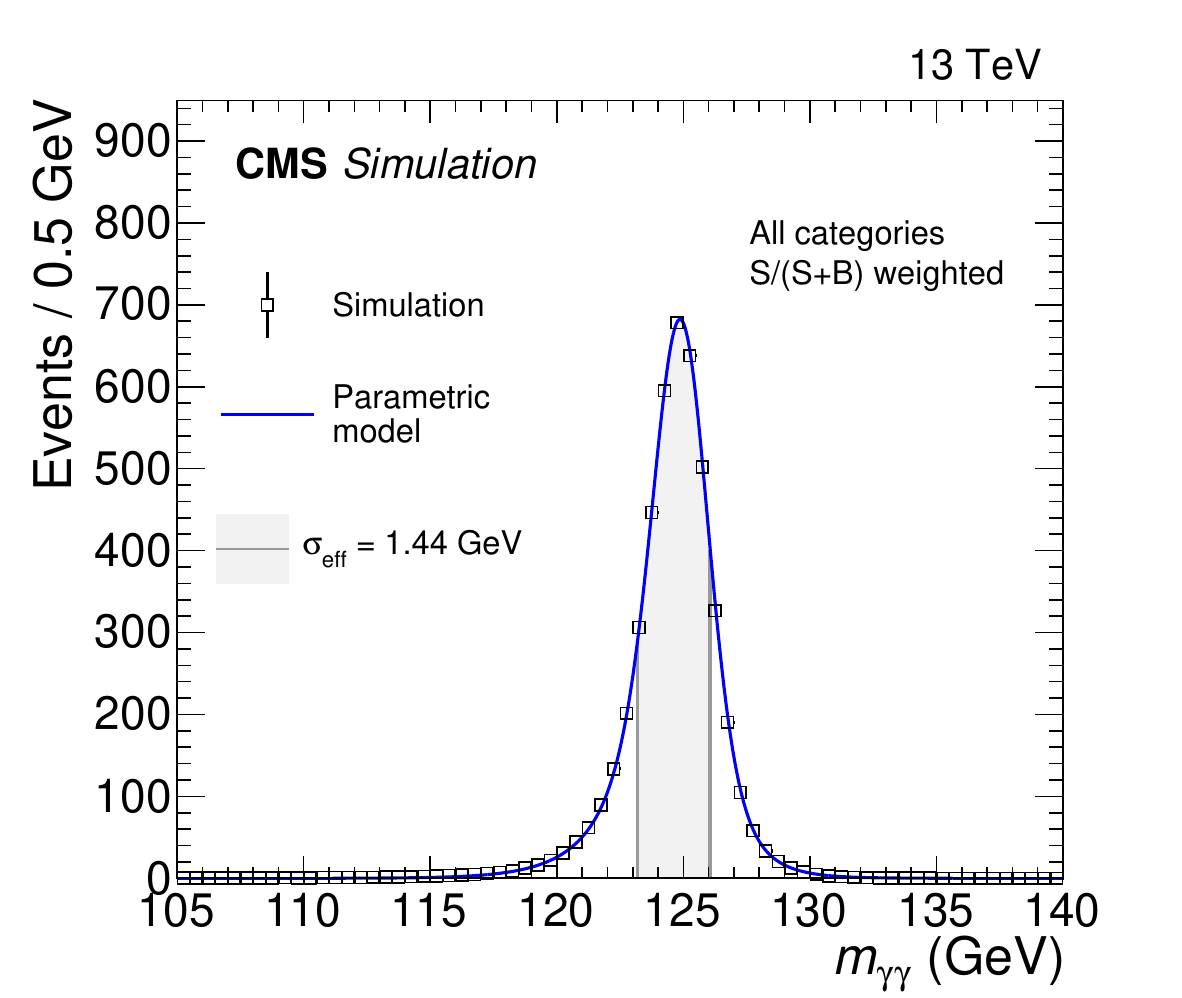}
  \caption{Signal model for the analysis category with the best mass resolution (\cmsLeft), and for all categories combined after scaling by their corresponding $\textrm{S/(S+B)}$ ratios (\cmsRight), for a simulated \hgg signal sample with $\mH = 125\GeV$. All Higgs boson production modes are summed.
}
  \label{fig:results_sigmodels}
\end{figure}

A simultaneous fit to signal samples at generated mass values of 120, 125 and 130\GeV is performed to determine the dependence of the parameters in the Voigtian functions on \mH. These are described by polynomials in \mH and are used in the signal model fit. The final fit function for each category is obtained by summing the functions for all production modes normalized to the expected signal yields in that category. Figure~\ref{fig:results_sigmodels} shows the signal model corresponding to $\mH = 125\GeV$ for the events with the highest diphoton BDT score, which populate the category with the best expected mass resolution. The open squares represent the weighted simulated events and the blue line the corresponding signal model. In addition, the effective standard deviation, $\sigma_\text{eff}$, defined as half of the smallest interval containing 68.3\% of the invariant mass distribution, is given. Also shown in the same figure is the signal model for the sum of all categories, with each category weighted by the corresponding $S/(S+B)$ ratio.

\subsection{Background model}
\label{sec:results_background}

The model used to describe the background for each of the analysis categories is obtained from data using the discrete profiling method, described in Ref.~\cite{DiscreteProfilingMethod}.
In this method, a large set of candidate function families are considered, including exponential functions, Bernstein polynomials, Laurent series, and power law functions. These functions are fitted to the \mgg distribution in the mass range of 100 to 180\GeV, excluding the signal region between 115 and 135\GeV.
For each family of functions, an F-test~\cite{fisher} is performed to determine the maximum order to be used in the fit. In addition, functions are retained for inclusion in the fit envelope if they satisfy a $\chi^2$ goodness-of-fit criterion corresponding to a probability greater than 0.01.

The choice of the background function is treated as a discrete nuisance parameter in the likelihood~\cite{Run1HGG} used to extract the best fit value of the Higgs boson mass, which accounts for the uncertainty associated with the arbitrary choice of the function.

\section{Systematic uncertainties}
\label{sec:systematics}

The systematic uncertainties are implemented as nuisance parameters in the likelihood function used to extract the best fit value of the Higgs boson mass. They are treated differently, depending on their effect on the signal model in the different analysis categories. The systematic uncertainties in the photon energy scale and resolution modify the shape of the signal model, while the other systematic uncertainties affect only the event yield.

Systematic uncertainties related to the energy scale are discussed in the following, whereas all other sources of uncertainty are described in Ref.~\cite{CMS:2021kom}.

\subsection{Uncertainties in the photon energy resolution correction}
\label{sec:smearingsystematics}

The uncertainties in the energy resolution corrections are evaluated using \Ztoee events.
\begin{itemize}

    \item \textit{Selection:} A loose selection on the electron ID is used for the nominal measurement. The associated uncertainty is estimated by applying a tighter selection on the electron ID and taking the difference with respect to the nominal result as a systematic uncertainty.

    \item \textit{Statistical uncertainty:} This uncertainty arises from the limited data and simulated sample sizes~\cite{IJazZ}. 

    \item \textit{Mass window size:} The uncertainty related to the choice of the invariant mass window used in the fit is evaluated as the difference between results obtained with two mass ranges, 75--100\GeV and 80--100\GeV, the latter being used for the final measurement. This systematic uncertainty accounts for possible biases in the extracted resolution corrections due to discrepancies in the tails of the \Ztoee mass distribution caused by mismodeling of background processes.

    \item \textit{Acceptance:} This uncertainty accounts for potential mismodeling of the trigger efficiency and the residual differences of data-to-simulation acceptance. It is estimated as the difference between the nominal corrections and the corrections derived without $\eta$ reweighting, while retaining all other reweighting corrections.

\end{itemize}

The uncertainties in the energy resolution corrections range from 5\% to 10\% and contribute a limited amount to the total systematic uncertainty, as expected. Their impact on the measurement is found to be $3\MeV$, as detailed in Table~\ref{tab:syst_per_source} of Section~\ref{subsec:impacts}.

\subsection{Uncertainties in the photon energy scale correction}

Several sources of systematic uncertainties in the photon energy scale derivation are evaluated using \Ztoee or \Ztomumug events. The former are treated as decorrelated across data taking periods, while the latter are treated as correlated. The sources of uncertainties considered in the measurement are:

\begin{itemize}
    
    \item \textit{Statistical uncertainty:} The uncertainties due to the limited number of \Ztomumug events in data and simulation are evaluated during the derivation of the photon energy scale corrections~\cite{IJazZ}. The total statistical uncertainty ranges from 0.05 to 0.2\%.
    
    \item \textit{Muon momentum scale uncertainty:} There are different sources of systematic uncertainty associated with the muon momentum scale calibration~\cite{Bodek:2012muon}, such as the statistical uncertainty of the event samples used to derive them; the uncertainty associated with the selection acceptance; the uncertainties arising from generator-level corrections and differences between generators. For each one, the energy scale corrections from \Ztomumug photons are rederived and compared with the nominal energy scale corrections. The resulting differences are then summed in quadrature. The total muon scale uncertainty ranges from 0.02 to 0.05\%.

    \item \textit{Residual non-linearity at high-\Et:} The \Ztomumug sample is statistically limited in the region with high-\Et photons, preventing the determination of reliable energy scale corrections in fine $\abs{\eta}$ bins. Instead, the \Et-dependent uncertainty due to the non-linearity of the photon energy response is estimated by fitting the residual energy scale as a function of \Et separately in the barrel and the endcaps. In both cases, the fitted dependence is consistent with a constant within the uncertainties. The magnitude of the fit uncertainty evaluated above 80\GeV in \Et is assigned as a systematic uncertainty. This results in a 0.15\% uncertainty assigned to photons in the barrel and in a 0.25\% uncertainty assigned to photons in the endcaps. For photons with energies less than 80\GeV the uncertainty is set to zero.

    \item \textit{Electromagnetic shower:} This uncertainty takes into account any possible bias in the energy scale corrections arising from the photon ID requirement, which relies on the electromagnetic shower shape variables, and is tighter than the one used for the derivation of photon energy scale corrections. The `electromagnetic shower' uncertainty is obtained from \Ztoee events due to their larger statistical precision and validated with \Ztomumug events. The photon ID score requirement is varied from its nominal value to a tighter value, corresponding to signal efficiencies of 95\% and 60\%, respectively. The difference in the energy scale correction is taken as the systematic uncertainty. We also consider additional subdominant sources of uncertainty described in Section~\ref{sec:smearingsystematics}, the `mass window size' and the `acceptance', which affect the energy scale. These are added in quadrature to the `electromagnetic shower' uncertainty. The resulting combined uncertainty is below 0.05\% in the region $\abs{\eta} < 1.0$ and between 0.1 and 0.3\% elsewhere.
    
    \item \textit{Energy resolution:} While deriving the final energy scale corrections of photons, a small bias of 0.017\% on the energy scale is observed when applying a selection on the steeply falling photon transverse energy spectrum in the presence of an additional 0.7\% energy resolution mismodeling, as shown in Ref.~\cite{IJazZ}. The 0.7\% corresponds to the impact of the average uncertainty of the resolution corrections to the corrected energy in simulated events.
    
    \item \textit{Selection:} When the photon and the muon are in close proximity, the muon energy deposition (${\approx}300\MeV$) can bias the photon cluster energy estimate. This effect is accurately modeled in simulation at the few-percent level and is therefore not expected to introduce a significant bias in the measurement. The dependence of the photon energy scale correction on $\Delta R(\PGm, \PGg)$ is studied to assess potential residual biases, which are found to be negligible, less than 0.02\%, and are therefore neglected. 
\end{itemize}

The contribution of each of these sources of uncertainty is shown in Fig.~\ref{fig:ZmmgScaleUnc} for photons in the barrel region.
\ifthenelse{\boolean{cms@external}}
{\begin{figure}[htb!]
\centering
\includegraphics[width=0.99\columnwidth]{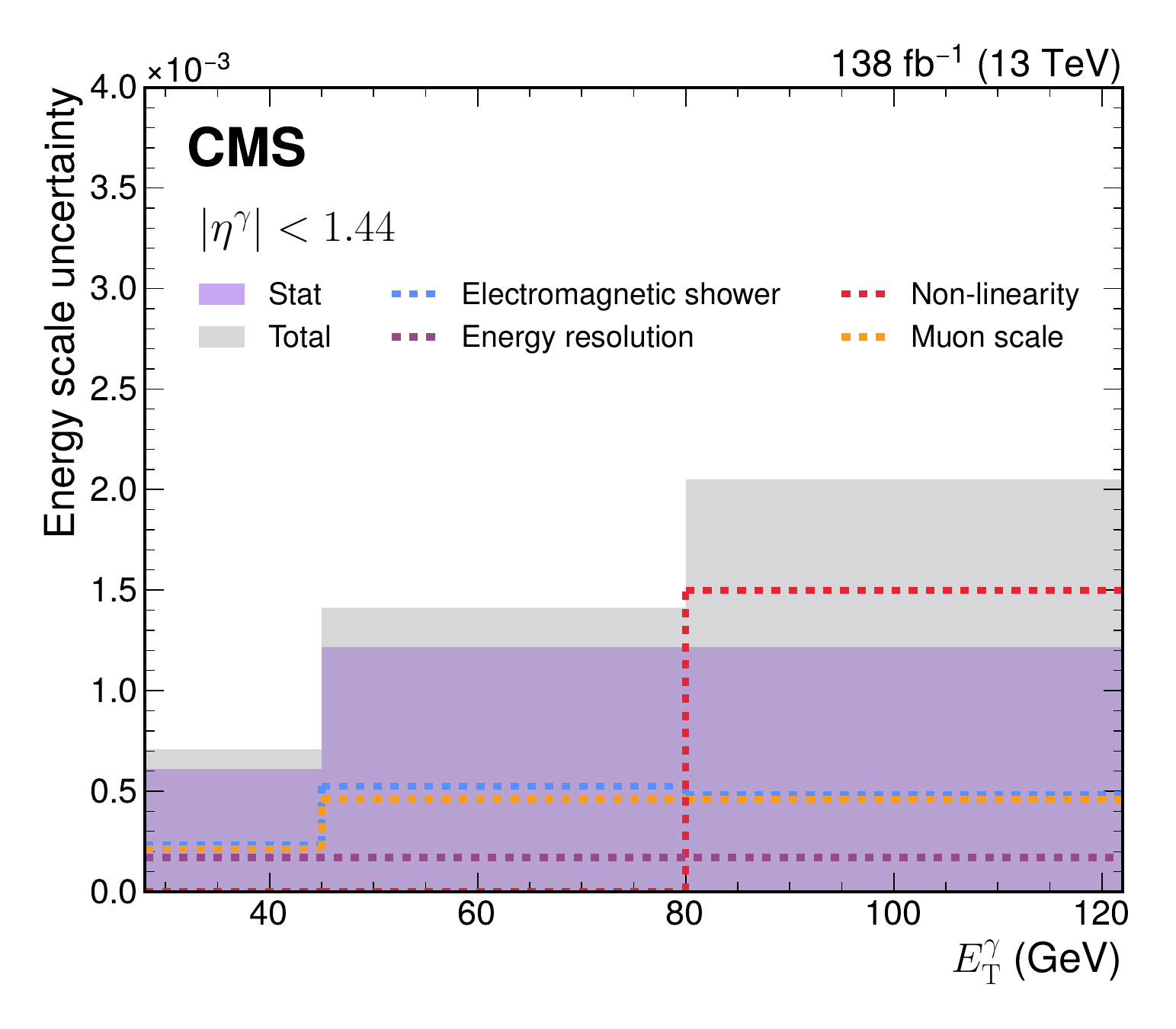}
\caption{Uncertainties in the energy scale as a function of $E^{\PGg}_{\mathrm{T}}$ of the photon in the barrel region, derived from \Ztomumug events. The shaded bands indicate the statistical and total uncertainties, while the dashed lines show the individual contributions from energy resolution, electromagnetic shower modeling, muon momentum scale, and non-linearity.}
\label{fig:ZmmgScaleUnc}
\end{figure}}
{\begin{figure}[htb!]
\centering
\includegraphics[width=0.55\textwidth]{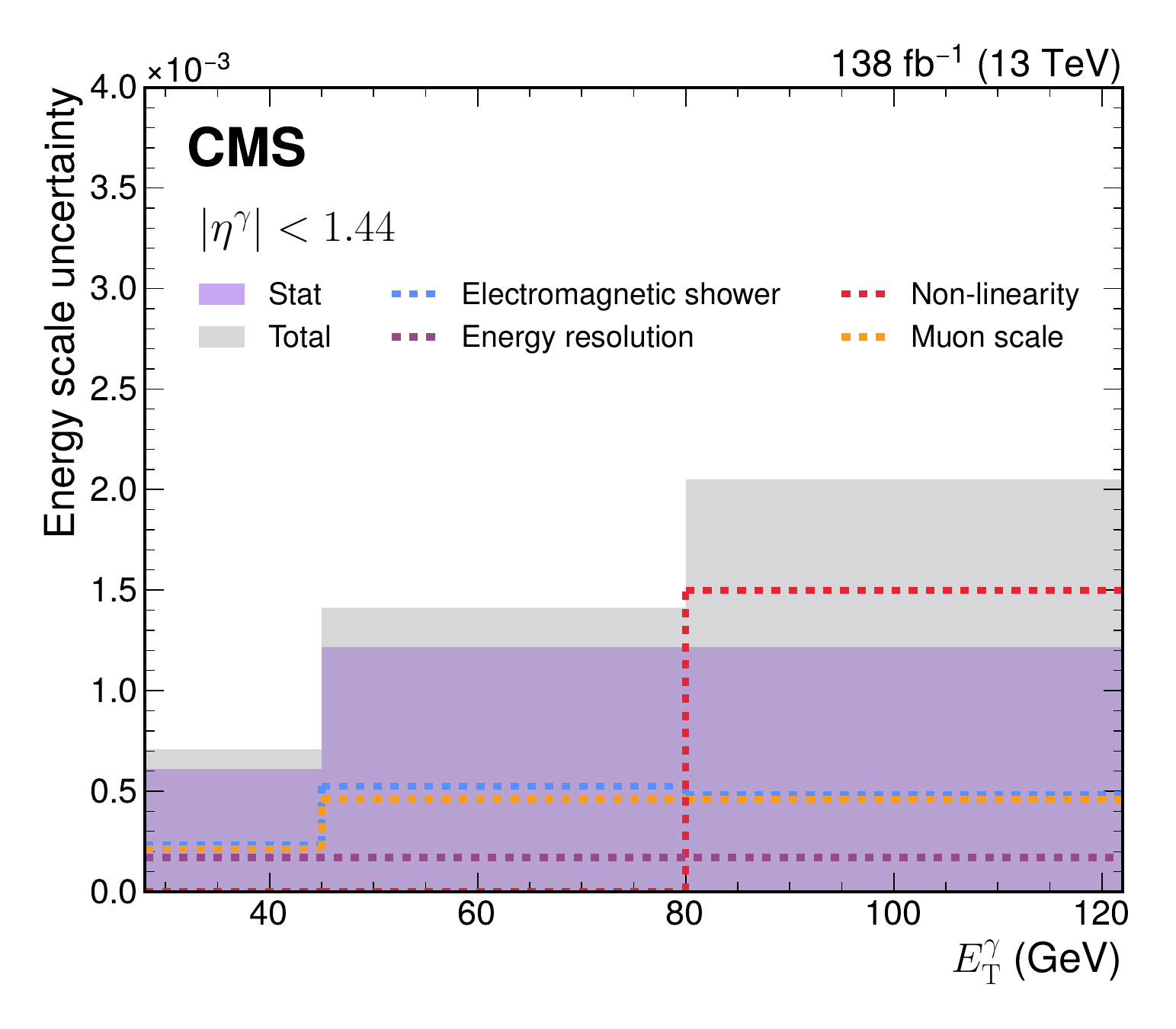}
\caption{Uncertainties in the energy scale as a function of $E^{\PGg}_{\mathrm{T}}$ of the photon in the barrel region, derived from \Ztomumug events. The shaded bands indicate the statistical and total uncertainties, while the dashed lines show the individual contributions from energy resolution, electromagnetic shower modeling, muon momentum scale, and non-linearity.}
\label{fig:ZmmgScaleUnc}
\end{figure}}

\subsection{Impact of the sources of uncertainty}
\label{subsec:impacts}
To quantify the impact of individual sources of uncertainty on the Higgs boson mass, nuisance parameters are grouped into disjoint sets. The parameters in each set are then fixed to their post-fit values, and the resulting reduction in the uncertainty is used to determine the contribution of that set to the total uncertainty in \mH.
The observed impact of individual sources of uncertainty is summarized in Table~\ref{tab:syst_per_source}. The leading systematic uncertainty arises from the limited number of events in the data and simulated \Ztomumug samples used to derive residual corrections to the photon energy scale. The next largest contributions arise from the uncertainties associated with the muon momentum scale calibration and with the extraction of the energy scale corrections for photons with $\Et > 80\GeV$ in the \Ztomumug sample. The systematic uncertainty associated with the interference between the gluon-fusion-induced \hgg signal process and the continuum diphoton background is estimated from the difference between fits using the standard signal simulation and a signal simulation that includes interference. The magnitude of the difference between the Higgs boson peak position in the two scenarios is found to be  
27\MeV and is assigned in its entirety as a systematic uncertainty. The 17\MeV contribution labeled `other sources' accounts for uncertainties related to the integrated luminosity measurement; simulation-to-data corrections applied to the signal; theoretical uncertainties in the signal prediction, including the branching fraction, the strong coupling constant $\alpS$, the renormalization and factorization scales, and the choice of parton distribution functions; the fraction of events for which the correct diphoton vertex is chosen; and variations of the inputs of the diphoton BDT classifier within their uncertainties, which could lead to event migrations between signal categories. Finally, when evaluating the statistical uncertainty, the constrained nuisance parameters associated with systematic uncertainties are fixed to their nominal values. The discrete nuisance parameters describing the choice of the background function, as well as the parameters describing the background shape for each candidate function, are unconstrained in the likelihood function. Consequently, they remain profiled in the fit, and the effect of the choice of the background function is included in the statistical component of the uncertainty.

    \begin{table}[htbp]
    \centering
   \topcaption{Observed impact of the different sources of uncertainty in the measurement of \mH}
    \label{tab:syst_per_source}
    \cmsTable{
    \begin{tabular}{l c}
    \hline
    Source & Contribution ({\MeVns}) \\
    \hline

    \multicolumn{2}{l}{Photon energy scale} \\
    \quad   Statistical uncertainty of muon sample& 74 \\
    \quad   Muon momentum & 55 \\
    \quad   Residual non-linearity & 54 \\
    \quad   Electromagnetic shower & 27 \\
    \quad   Energy resolution & 21 \\

    Photon energy resolution & 3 \\
    Interference between $\Pg\Pg\PH$ signal and background & 27 \\ 
    Other sources & 17 \\
    \hline
    Systematic uncertainty & 117 \\
    Statistical uncertainty & 98 \\
    Total uncertainty & 153 \\
    \hline
    \end{tabular}
    }
    \end{table}

\section{Results}
\label{sec:results}
\begin{figure}[htb!]
    \centering
    \includegraphics[width=0.49\textwidth]{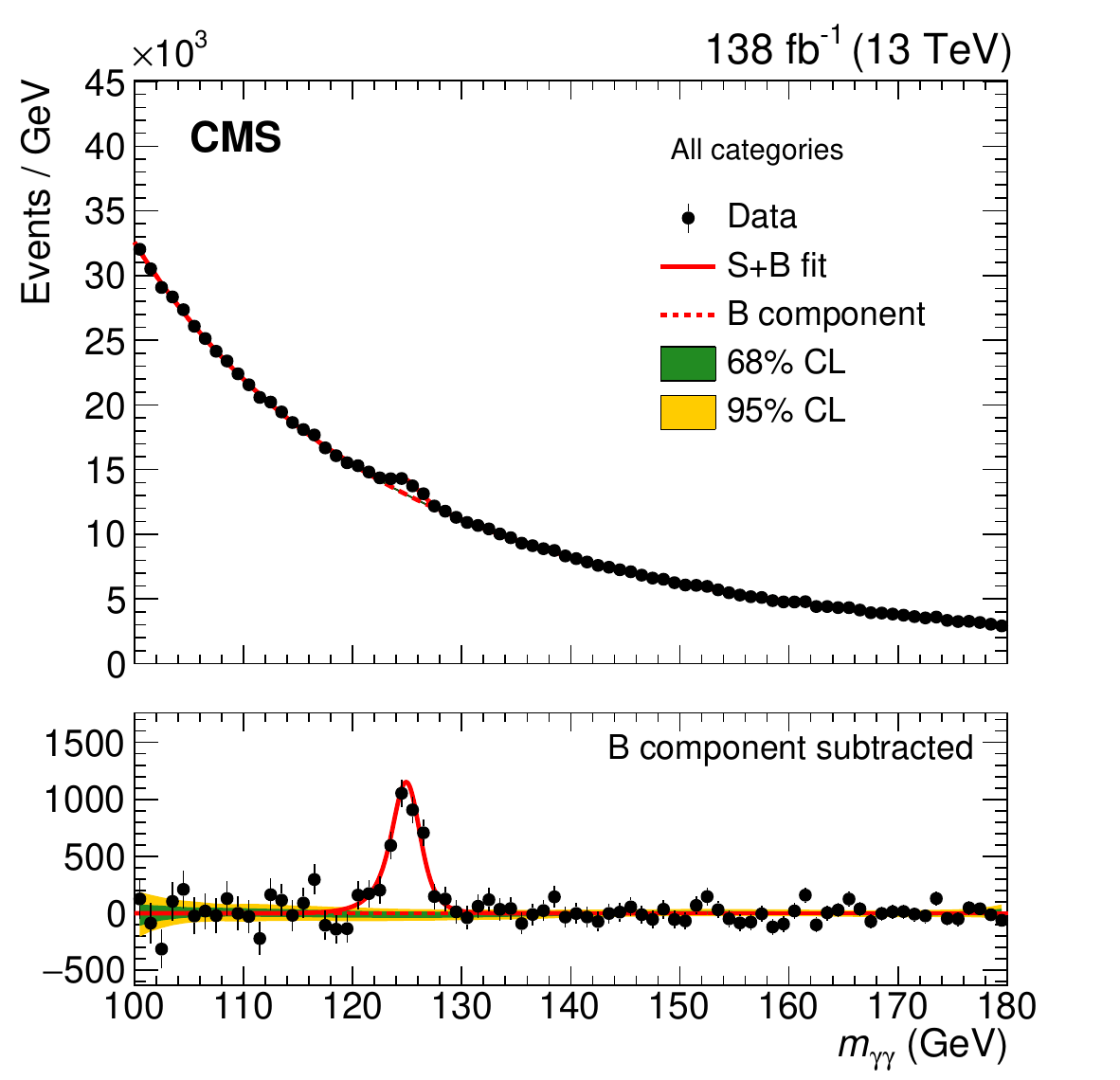}
    \includegraphics[width=0.49\textwidth]{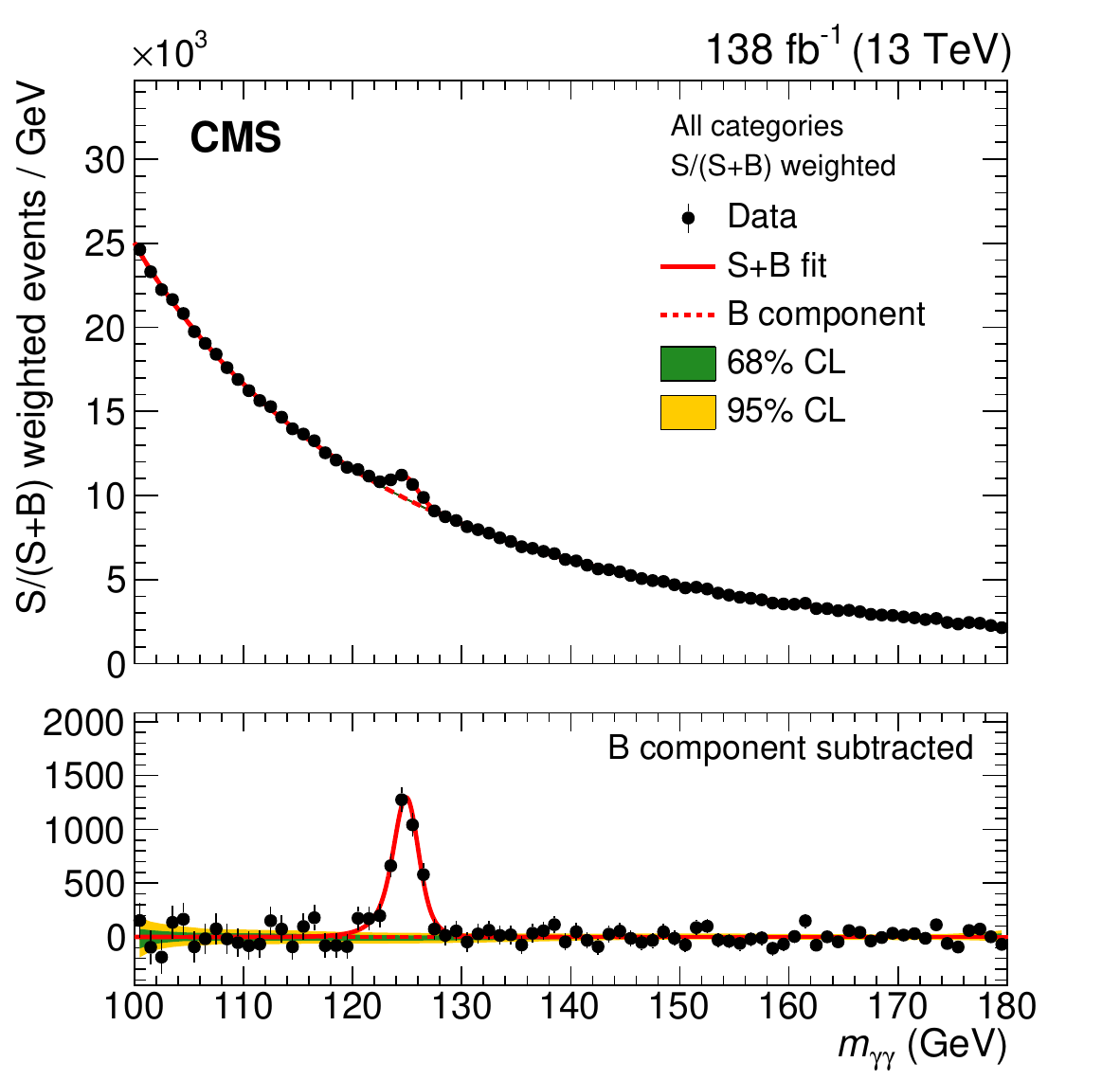}
    \caption{Data and combined signal and background model fit for all analysis categories, unweighted (\cmsLeft) and weighted by their sensitivity (\cmsRight). The one (green) and
two (yellow) standard deviation bands include the uncertainties in the background component of the fit. The lower panel shows the residuals after the background is subtracted.}
    \label{fig:SplusBFits}
\end{figure}

The results are obtained with an asymptotic approach~\cite{Asymtotic1,Asymtotic2}, using a test statistic based on the profile likelihood ratio~\cite{TestStatistic} as implemented in the CMS statistical analysis tool \textsc{Combine}~\cite{CMS:2024onh}. In the fit, the signal strength, $\mu = \sigma / \sigma_\mathrm{SM}$, is allowed to vary, while \gammah is fixed to 4.1\MeV. The data and the combined signal-plus-background model for all categories, both unweighted and weighted by their $S/(S+B)$ ratios, are shown in Fig.~\ref{fig:SplusBFits}. By integrating the post-fit signal-plus-background and background-only models between 115 and 135\GeV, the signal yield is found to be approximately 4320 events.

The best fit value of the Higgs boson mass is found to be: 
\ifthenelse{\boolean{cms@external}}
{\begin{equation*}
\begin{aligned}
\mH & = 125.13 \pm 0.15\GeV \\
& = 125.13 \pm 0.10\stat \pm 0.12\syst\GeV.
\end{aligned}
\end{equation*}}
{\begin{equation*}
\mH = 125.13 \pm 0.15\GeV = 125.13 \pm 0.10\stat \pm 0.12\syst\GeV.
\end{equation*}}

\begin{figure}[htb!]
    \centering
    \includegraphics[width=\cmsFigWidth]{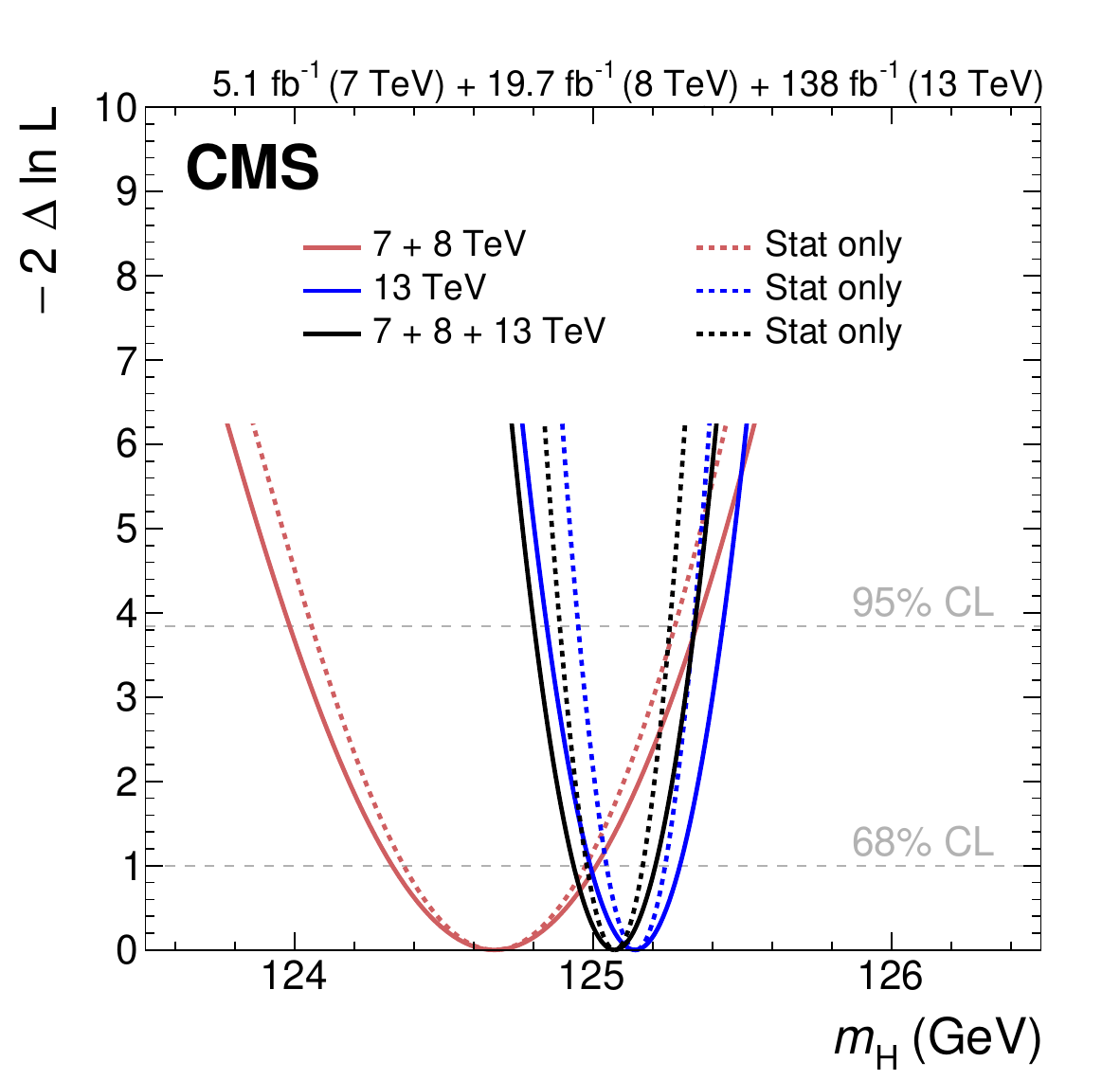}
	 \caption{Likelihood scans of the Higgs boson mass measured in the \hgg decay channel for the $\sqrt{s} = 7$ and $8\TeV$ data sets, the $\sqrt{s} = 13\TeV$ data set, and their combination. Solid lines show the full likelihood scan including systematic uncertainties, while dashed lines correspond to the statistical-only case.}
    \label{fig:money_massNwidth_scan}
\end{figure}

\begin{figure*}[htb!]
    \centering
    \includegraphics[width=0.8\textwidth]{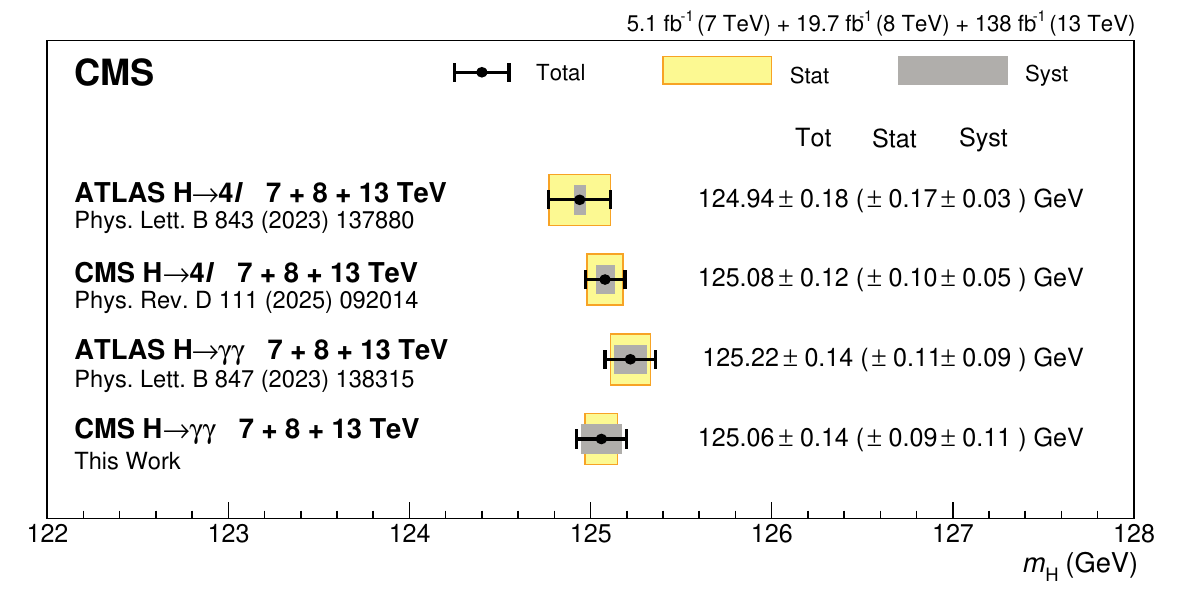}
	 \caption{Summary of the ATLAS and CMS Higgs boson mass measurements using the diphoton and the four-lepton final states, combining results at $\sqrt{s} = 7$, 8 and 13\TeV. The black points represent the best fit values of each measurement. The yellow and gray bands show the statistical and systematic uncertainties in each measurement, respectively. The horizontal black bars show the total uncertainties. The value of each measurement is given, along with the total uncertainties, splitting statistical only and systematic only uncertainties in parentheses.}
    \label{fig:atlas_cms_summary_mass}
\end{figure*}

For comparison, the expected uncertainties are $\pm 0.10\stat \pm 0.12\syst\GeV$, evaluated using an Asimov data set~\cite{TestStatistic} generated from the expected SM signal with the best fit background model. A scan of twice the negative logarithm of the likelihood as a function of \mH is shown in Fig.~\ref{fig:money_massNwidth_scan}. In addition, a measurement of the signal strength parameter $\mu$ is performed using the full data set, allowing the \mH parameter to vary in the fit. The best fit value is found to be $\mu = 0.95 \pm 0.06\stat \pm 0.07\syst$. This result agrees with the dedicated measurement of Higgs boson properties in the diphoton channel with the same data set~\cite{CMS:2021kom}, and is compatible with the standard model prediction, $\mu = 1$. 
This mass measurement, obtained with CMS 13\TeV data, is found to be compatible with the previous CMS measurement combining 7 and 8 TeV samples at the level of 1.2 standard deviations. The two measurements are combined assuming no correlations, as both the recorded and simulated data sets are independent. The result is shown in Fig.~\ref{fig:money_massNwidth_scan}. The combination gives:
\ifthenelse{\boolean{cms@external}}
{\begin{equation*}
\begin{aligned}
\mH & = 125.06 \pm 0.14\GeV  \\
& =125.06 \pm 0.09\stat \pm 0.11 \syst\GeV.
\end{aligned}
\end{equation*}}
{\begin{equation*}
\mH = 125.06 \pm 0.14\GeV = 125.06 \pm 0.09\stat \pm 0.11 \syst\GeV.
\end{equation*}}
A summary of the ATLAS and CMS Higgs boson mass measurements using the diphoton and the four-lepton final states, combining results at $\sqrt{s} = 7$, 8, and 13\TeV, is shown in Fig.~\ref{fig:atlas_cms_summary_mass}.

Numerical results of the analysis are available in HEPData~\cite{HEPData}.

\section{Summary}
\label{sec:summary}

A new measurement of the Higgs boson mass has been conducted in the diphoton decay channel, using the data set collected by CMS between 2016 and 2018 at $\sqrt{s} = 13\TeV$ at the CERN LHC. New analysis techniques, enabled by the increased integrated luminosity, were used to improve the measurement precision, refine the detector calibration, and derive corrections accounting for differences between photons and electrons, resulting in a nearly factor-of-two reduction in the systematic uncertainty compared to previous measurements by the CMS Collaboration.

The main improvements with respect to the previous analysis~\cite{higgs_mass} include the derivation of granular \Et-dependent energy resolution corrections, a new simulation-based method to correct for differences between the photon and electron energy scales due to radiation damage in the ECAL crystals, and a calibration procedure using \Ztomumug events to correct residual energy scale differences between electrons and photons.
In addition, a novel signal-to-background classifier was employed, with backgrounds containing at least one jet misidentified as a photon estimated using a technique based on control samples in data. To further enhance the measurement precision, multiple analysis categories were defined according to the classifier output, using a figure of merit that accounts for both the signal-to-background ratio and the relative diphoton mass resolution.

The Higgs boson mass is measured to be $\mH = 125.13 \pm 0.15\GeV = 125.13 \pm 0.10\stat \pm 0.12\syst\GeV$ using data collected at $\sqrt{s} = 13\TeV$. When combined with the corresponding measurement from CMS data at $\sqrt{s} = 7$ and 8\TeV, the mass is determined to be $\mH = 125.06 \pm 0.14\GeV = 125.06 \pm 0.09\stat \pm 0.11\syst\GeV$. 

\begin{acknowledgments}
    We congratulate our colleagues in the CERN accelerator departments for the excellent performance of the LHC and thank the technical and administrative staffs at CERN and at other CMS institutes for their contributions to the success of the CMS effort. In addition, we gratefully acknowledge the computing centers and personnel of the Worldwide LHC Computing Grid and other centers for delivering so effectively the computing infrastructure essential to our analyses. Finally, we acknowledge the enduring support for the construction and operation of the LHC, the CMS detector, and the supporting computing infrastructure provided by the following funding agencies: SC (Armenia), BMFWF and FWF (Austria); FNRS and FWO (Belgium); CNPq, CAPES, FAPERJ, FAPERGS, and FAPESP (Brazil); MES and BNSF (Bulgaria); CERN; CAS, MoST, and NSFC (China); MINCIENCIAS (Colombia); MSES and CSF (Croatia); RIF (Cyprus); SENESCYT (Ecuador); ERC PRG and PSG, TARISTU24-TK10 and MoER TK202 (Estonia); Academy of Finland, MEC, and HIP (Finland); CEA and CNRS/IN2P3 (France); SRNSF (Georgia); BMFTR, DFG, and HGF (Germany); GSRI (Greece); MATE and NKFIH (Hungary); DAE and DST (India); IPM (Iran); SFI (Ireland); INFN (Italy); MSIT and NRF (Republic of Korea); MES (Latvia); LMTLT (Lithuania); MOE and UM (Malaysia); BUAP, CINVESTAV, CONACYT, LNS, SEP, and UASLP-FAI (Mexico); MOS (Montenegro); MBIE (New Zealand); PAEC (Pakistan); MSHE, NSC, and NAWA (Poland); FCT (Portugal); MESTD (Serbia); MICIU/AEI and PCTI (Spain); MOSTR (Sri Lanka); Swiss Funding Agencies (Switzerland); MST (Taipei); MHESI (Thailand); TUBITAK and TENMAK (T\"{u}rkiye); NASU (Ukraine); STFC (United Kingdom); DOE and NSF (USA).

\begin{sloppypar}
\setlength\emergencystretch{\hsize}
\hyphenation{Rachada-pisek} Individuals have received support from the Marie-Curie program and the European Research Council and Horizon 2020 Grant, contract Nos.\ 675440, 724704, 752730, 758316, 765710, 824093, 101115353, 101002207, 101001205, and COST Action CA16108 (European Union); the Leventis Foundation; the Alfred P.\ Sloan Foundation; the Alexander von Humboldt Foundation; the Science Committee, project no. 22rl-037 (Armenia); the Fonds pour la Formation \`a la Recherche dans l'Industrie et dans l'Agriculture (FRIA) and Fonds voor Wetenschappelijk Onderzoek contract No. 1228724N (Belgium); the Beijing Municipal Science \& Technology Commission, No. Z191100007219010, the Fundamental Research Funds for the Central Universities, the Ministry of Science and Technology of China under Grant No. 2023YFA1605804, the Natural Science Foundation of China under Grant No. 12535004, and USTC Research Funds of the Double First-Class Initiative No.\ YD2030002017 (China); the Ministry of Education, Youth and Sports (MEYS) of the Czech Republic; the Shota Rustaveli National Science Foundation (Georgia); the Deutsche Forschungsgemeinschaft (DFG), among others, under Germany's Excellence Strategy -- EXC 2121 ``Quantum Universe" -- 390833306, and under project number 400140256 - GRK2497; the Hellenic Foundation for Research and Innovation (HFRI), Project Number 2288 (Greece); the Hungarian Academy of Sciences, the New National Excellence Program - \'UNKP, the NKFIH research grants K 131991, K 138136, K 143460, K 143477, K 147557, K 146913, K 146914, K 147048, TKP2021-NKTA-64, and 2025-1.1.5-NEMZ\_KI-2025-00004, and MATE KKP and KKPCs Research Excellence and Flagship Research Groups grants (Hungary); the Council of Science and Industrial Research, India; ICSC -- National Research Center for High Performance Computing, Big Data and Quantum Computing, FAIR -- Future Artificial Intelligence Research, and CUP I53D23001070006 (Mission 4 Component 1), funded by the NextGenerationEU program, the Italian Ministry of University and Research (MUR) under Bando PRIN 2022 -- CUP I53C24002390006, PRIN PRIMULA 2022RBYK7T (Italy); the Latvian Council of Science; the Ministry of Science and Higher Education, project no. 2022/WK/14, and the National Science Center, contracts Opus 2021/41/B/ST2/01369, 2021/43/B/ST2/01552, 2023/49/B/ST2/03273, and the NAWA contract BPN/PPO/2021/1/00011 (Poland); the Funda\c{c}\~ao para a Ci\^encia e a Tecnologia (Portugal); the National Priorities Research Program by Qatar National Research Fund; MICIU/AEI/10.13039/501100011033, ERDF/EU, ``European Union NextGenerationEU/PRTR", projects PID2022-142604OB-C21, PID2022-139519OB-C21, PID2023-147706NB-I00, PID2023-148896NB-I00, PID2023-146983NB-I00, PID2023-147115NB-I00, PID2023-148418NB-C41, PID2023-148418NB-C42, PID2023-148418NB-C43, PID2023-148418NB-C44, PID2024-158190NB-C22, RYC2021-033305-I, RYC2024-048719-I, CNS2023-144781, CNS2024-154769 and Plan de Ciencia, Tecnolog{\'i}a e Innovaci{\'o}n de Asturias, Spain; the Chulalongkorn Academic into Its 2nd Century Project Advancement Project, the National Science, Research and Innovation Fund program IND\_FF\_68\_369\_2300\_097, and the Program Management Unit for Human Resources \& Institutional Development, Research and Innovation, grant B39G680009 (Thailand); the Eric \& Wendy Schmidt Fund for Strategic Innovation through the CERN Next Generation Triggers project under grant agreement number SIF-2023-004; the Kavli Foundation; the Nvidia Corporation; the SuperMicro Corporation; the Welch Foundation, contract C-1845; and the Weston Havens Foundation (USA).
\end{sloppypar}
\end{acknowledgments}
\bibliography{auto_generated}
\cleardoublepage \appendix\section{The CMS Collaboration \label{app:collab}}\begin{sloppypar}\hyphenpenalty=5000\widowpenalty=500\clubpenalty=5000\cmsinstitute{Yerevan Physics Institute, Yerevan, Armenia}
{\tolerance=6000
A.~Gevorgyan\cmsorcid{0000-0003-2751-9489}, A.~Hayrapetyan, V.~Makarenko\cmsorcid{0000-0002-8406-8605}, A.~Tumasyan\cmsAuthorMark{1}\cmsorcid{0009-0000-0684-6742}
\par}
\cmsinstitute{Institut f\"{u}r Hochenergiephysik, Vienna, Austria}
{\tolerance=6000
P.S.~Hussain\cmsorcid{0000-0002-4825-5278}, M.~Sonawane\cmsorcid{0000-0003-0510-7010}
\par}
\cmsinstitute{Marietta Blau Institute for Particle Physics, Vienna, Austria}
{\tolerance=6000
W.~Adam\cmsorcid{0000-0001-9099-4341}, L.~Benato\cmsorcid{0000-0001-5135-7489}, T.~Bergauer\cmsorcid{0000-0002-5786-0293}, M.~Dragicevic\cmsorcid{0000-0003-1967-6783}, S.~Gundacker\cmsorcid{0000-0003-2087-3266}, A.K.~Guven\cmsorcid{0009-0004-5670-5138}, M.~Jeitler\cmsAuthorMark{2}\cmsorcid{0000-0002-5141-9560}, N.~Krammer\cmsorcid{0000-0002-0548-0985}, A.~Li\cmsorcid{0000-0002-4547-116X}, D.~Liko\cmsorcid{0000-0002-3380-473X}, M.~Matthewman, A.~Pfeiffer\cmsorcid{0000-0001-5328-448X}, J.~Schieck\cmsAuthorMark{2}\cmsorcid{0000-0002-1058-8093}, R.~Sch\"{o}fbeck\cmsAuthorMark{2}\cmsorcid{0000-0002-2332-8784}, M.~Shooshtari\cmsorcid{0009-0004-8882-4887}, N.~Van~Den~Bossche\cmsorcid{0000-0003-2973-4991}, W.~Waltenberger\cmsorcid{0000-0002-6215-7228}, C.E.~Wulz\cmsAuthorMark{2}\cmsorcid{0000-0001-9226-5812}
% ...
\par}
$^{1}$Also at Yerevan State University, Yerevan, Armenia\\
$^{2}$Also at Technische Universit\"{a}t Wien, Vienna, Austria\\
% ...
\end{sloppypar}
%%% END EDITABLE REGION %%%
% skeleton_end
\end{document}